\documentclass[twocolumn,pre,aps,showpacs]{revtex4}

\usepackage{dcolumn,graphicx,amsmath,amssymb,pxfonts}

\newcommand{\mfr}{M_\mathrm{fr}}

\begin{document}

\title{On network bipartivity}

\author{Petter \surname{Holme}}
\email{holme@tp.umu.se}
\affiliation{Department of Physics, Ume{\aa} University, 
  901~87 Ume{\aa}, Sweden}

\author{Fredrik \surname{Liljeros}}
\affiliation{Department of Epidemiology, Swedish Institute for
  Infectious Disease Control, 171~82 Solna, Sweden}
\affiliation{Department of Sociology, Stockholm University, 106~91
  Stockholm, Sweden}

\author{Christofer R.\ \surname{Edling}}
\affiliation{Department of Sociology, Stockholm University, 106~91
  Stockholm, Sweden}

\author{Beom Jun \surname{Kim}}
\affiliation{Department of Molecular Science
  and Technology, Ajou University, Suwon 442-749, Korea}

\begin{abstract}
  Systems with two types of agents with a preference for heterophilous
  interaction produces networks that are more or less close to
  bipartite. We propose two measures quantifying the notion of
  bipartivity. The two measures---one well-known and natural, but
  computationally intractable; one computationally less complex, but
  also less intuitive---are examined on model networks that
  continuously interpolates between bipartite graphs and graphs with
  many odd circuits. We find that the bipartivity measures increase
  as we tune the control parameters of the test networks to
  intuitively increase the bipartivity, and thus conclude that the
  measures are quite relevant. We also measure and discuss the values
  of our bipartivity measures for empirical social networks
  (constructed from professional collaborations, Internet communities
  and field surveys). Here we find, as expected, that networks arising
  from romantic online interaction have high, and professional
  collaboration networks have low bipartivity values. In some other
  cases, probably due to low average degree of the network, the
  bipartivity measures cannot distinguish between romantic and
  friendship oriented interaction.
\end{abstract}


\maketitle

\section{Introduction\label{sec:intro}}

Any system, natural or man-made, consisting of entities that interact
pairwise can be described in terms of a network. Networks in the real
life often contain some degree of randomness, and has also some
structure arising from the strategies or laws the entities follow to
make new contacts. Such networks---that can only be described as having
both randomness and structure---are called complex networks and has
lately received much attention in the physicist
community~\cite{review,review2}. Among the most important developments
in this recent surge of activity in network research is arguably
the categorization and quantification of static network structures
such as clustering~\cite{WS}, degree distribution~\cite{sf},
assortative mixing coefficient~\cite{assmix}, grid
coefficient~\cite{grid}, etc. A network with no
circuit of odd length is called \textit{bipartite}. Many systems are
naturally modeled as bipartite networks: Biochemical networks can be
described by vertices representing chemical substances separated by
vertices representing chemical reactions~\cite{jeong}. As another
example, we have the so called ``two-mode'' representation of
affiliation networks where one kind of vertices represents e.g.\
organizations and the other type represents individual actors, and the
edges indicates to which organizations an actor belongs. But there are
also networks that are not necessarily bipartite, but closer to
bipartite than what can be expected from a completely random
network. Examples of such networks are those that are formed by two
types of agents with a preference for heterophilous interaction (human
sexual contacts~\cite{liljeros,lea}, and human romance or partnership
networks~\cite{partner} being two cases). In many cases one knows the
type of the individual vertices (the gender of the actors in the
examples above)~\cite{freeman}, but in other cases such information
might be lacking (the data studied in Ref.~\cite{HEL} for a concrete
example). Nevertheless, the `bipartivity'---how far away from being
bipartite a graph is---is a measurable structure; and therefore, we
believe, deserves attention.

How can we measure bipartivity? The idea we use in this paper is the
following: We suppose that all agents of one type tried their best in
forming a connection to an agent of the other type. Then we measure to
what extent this assumption fail. We can assign a label
$\sigma_v\in\{-1,+1\}$ to each vertex $v$ and check for the maximal
fraction of edges between vertices of different sign. This fraction
will be equal to or higher than the actual fraction of edges between
vertices of different type. But, at least for strong heterophilous
preference in the network formation, the difference should be
small. For weak heterophilous preference this approach will likely
fail to produce a correct classification of the individual vertices.
Still, the number of even circuits should be larger than in a
network created under the same circumstances but with no heterophilous
preference; and this will (as we will see) give a lower value of such
a bipartivity measure. So even if we cannot reproduce the correct
fraction of vertices of different type, we have a measure that is a
monotonous function of the strength of the heterophilous
preference. It is convenient (at least for people familiar with
statistical mechanics) to phrase a problem like this in terms of the
antiferromagnetic Ising model. Our bipartivity measure---the maximal
fraction of edges between vertices of different sign---is directly
related to the ground state energy of the antiferromagnetic Ising
model (the relation is given in Sect.~\ref{sec:b1def}). Throughout the
paper we will often use the terminology of such spin systems, such as
the antiferromagnetic Ising model. For example we talk of an edge
between two vertices of the same tag as a `frustrated' edge.

The spin system analogy to combinatorial optimization problems such
as the one we are facing---to find minimal fraction of frustrated
edges---is nothing new. With this approach the fraction of frustrated
edges defines a cost function corresponding to the energy of the spin
system. The two most studied problems in this area are the
$p$-coloring problem and the graph bisection problem. In the
$p$-coloring problem the question is whether or not  the vertices of a
graph can be assigned one of $p$ colors in such a way that no edge
goes between two vertices of the same color. This problem is solvable
in linear time for $p=2$, but NP-complete (i.e.\ in the general case
not calculable in polynomial time~\cite{hope}) for $p>2$. The graph
bisection problem (also NP-complete) is to partition the vertex-set
into two sets of equal size such that the number of edges between the
two sets is minimized~\cite{jerrum,schreiber,FA}. Both these problems
can, just as ours, be phrased in terms of spin-models with
antiferromagnetic interaction. Our minimization problem is a little
bit different from the bisection problem in that the two sections can
have arbitrary sizes. However, as in the bisection and $p$-coloring
problems we are also faced with an NP-complete optimization
problem. (Our aim---to find the ground state energy of
antiferromagnetic Ising model can be mapped to a min-flow max-cut
problem~\cite{alava} which is NP-hard on general
networks~\cite{karp}.)

As the spin models of statistical physics are familiar to statistical
physicists, it is not surprising that topics like the Ising and
\textsl{XY} models on various model networks~\cite{bw,spin} have received
much attention in physicists' network literature. The motivation for
such studies, as models of real-world systems, is that they can capture
some features of opinion formation or similar social
processes~\cite{socstatmech}. The present work can also be described
as a study of a spin model on a complex network, but unlike the above
mentioned studies, the spin model is used as a tool to measure a
static network structure.

\section{The measures}
In the following sections we will go through the two bipartivity
measures. We state the definitions, dissect the algorithms and give
analytic discussions about the limit properties.

We represent a undirected network by $G=(V,E)$ and a directed network by
$G_\mathrm{dir}=(V,A)$, where $V$ is the set of vertices, $E$ is a set
of edges (or undirected pairs of vertices), and $A$ is a set of arcs
(or ordered pairs of vertices). A \textit{path} of length $l$ is a
sequence of vertices $v_1,\cdots,v_l$ such that $(v_i,v_{i+1})\in E$
(or $(v_i,v_{i+1})\in A$ for directed graphs); a \textit{circuit} is a
path where the first and last vertex are identical. In an
\textit{elementary} path, or circuit, no vertex appears twice (except
the first and last in case of circuits). In the present paper we will
only talk about elementary paths and circuits---so, for brevity we omit
the word `elementary.' Throughout the paper, when necessary, we let
sub- or superscript `dir' denote directed versions of quantities. In
many cases the generalization from undirected to directed networks is
straightforward; in these cases we will pursue the discussion in the
framework of undirected networks.

\subsection{The measure $b_1$}
\subsubsection{Definition\label{sec:b1def}}
The first measure we consider is simply the fraction of unfrustrated
edges in the ground state of the antiferromagnetic Ising model on the
network. In terms of the antiferromagnetic Ising model the quantity
can be written as
\begin{equation}
  b_1 = 1-\frac{\mfr}{M}=\frac{1}{2}-\frac{E_0}{2M}~,\label{eq:b1}
\end{equation}
where $\mfr$ is the number of frustrated edges in the ground state
(the usual cost function in the two-coloring problem).
$E_0$ is the ground state energy
\begin{equation}
  E_0=\min_{\{\sigma_v\}}~, H\label{eq:e0}
\end{equation}
where $H$ is the Hamiltonian of the antiferromagnetic Ising model:
\begin{subequations}
\begin{eqnarray}
H&=&\sum_{(v,w)\in  E} \sigma_v\sigma_w\\
H_\mathrm{dir}&=&\sum_{(v,w)\in  A} \sigma_v\sigma_w
\end{eqnarray}
\end{subequations}
The directed quantity is obtained by substituting $H$ by
$H_\mathrm{dir}$ in Eqs.~(\ref{eq:b1}) and (\ref{eq:e0}), and edges by
arcs in the above discussion. The topology of the energy landscape is
determined by the underlying network, and can in general be very
complex~\cite{barahona}.

\subsubsection{Limit properties}

The $b_1$ measure takes values in the interval $(1/2,1]$. The upper
bound is attained for bipartite graphs. It is easy to see that $b_1$
cannot be lower than $1/2$: Consider a ground state configuration for
which the opposite is true. Then there must be at least one vertex
with more than half of its edges frustrated. Flipping this spin
would reduce the energy, which contradicts the fact that the system is
in the ground state. We do not know if this bound is realized for any
finite graphs, but $b_1=1/2$ is the limit value for $b_1$ for a fully
connected graph as $N\rightarrow\infty$: Partition the fully connected
graph $K_N$ of $N$ vertices (and $M=N(N-1)/2$ edges) into one set of
$N'$ and one set of $N-N'$ vertices and assign opposite spins to the
elements of these sets. The number of frustrated edges is precisely
the number of edges within each set which is:
\begin{eqnarray}
  \mfr(K_N)&=&\frac{N'(N'-1)}{2}+\frac{(N-N')(N-1-N')}{2}
  \nonumber\\&=&M-N'(N-N')~.
\end{eqnarray}
Thus the minimum number of frustrated edges is exactly $N^2/4-N/2$ for
$N'=N/2$, and the fraction of unfrustrated edges is
\begin{equation}
  b_1 = \frac{1}{2-2/N}\rightarrow\frac{1}{2}
  \mbox{~as~} N\rightarrow\infty~.
\end{equation}
The above arguments can be generalized to directed networks
straightforwardly.

\subsubsection{Minimization by exchange Monte Carlo}
The complexity of the ``energy landscape'' of the antiferromagnetic
Ising model on an arbitrary network is difficult to judge \textit{a
  priori}. There are indications that no natural network would be
too hard for a regular simulated annealing
approach~\cite{simann,jerrum}. To be on safer ground, we use a Monte Carlo
scheme that is evidently very efficient to sweep even an extremely
`rugged' energy landscape without getting stuck in local minima---the
so called exchange Monte Carlo (XMC)~\cite{xmc}. The idea of exchange
Monte Carlo is to run standard Metropolis Monte Carlo for $N_T$
replicas of the system, each at a specific temperature. Then from time
to time two replicas at adjacent temperatures are compared, and with a
probability\begin{equation}
  P_\mathrm{exch.}=\left\{\begin{array}{ll} 1 & \mbox{if $\Delta<0$}\\
      e^{-\Delta} & \mbox{otherwise}\end{array}\right.~,
\end{equation}
where
\begin{equation}
   \Delta=\left(\frac{1}{T}-\frac{1}{T'}\right)(E'-E)~,
\end{equation}
and $E$ is the energy of the configuration at temperature $T$
(similarly for $T'$ and $E'$), and $T<T'$.
the two replicas are swapped between the temperatures. This condition
is designed so that the Monte Carlo scheme preserves the Boltzmann
distribution. This is not decisive for us who are looking for the
ground state energy, rather that performing a proper sampling of the
configuration space, but anyway kept in our measurements. Besides just
running the XMC scheme we also periodically quench the system,
i.e.\ we sweep through all vertices of the network consecutively and flip
spins that lower the energy. The sweeps are continued until a sweep
with no spin-flips has occurred. For later reference we introduce the
notations $t_\mathrm{avg}$ for the total number of MC sweeps---we
refer to the number of MC sweeps as `time'---$t_\mathrm{quench}$ for
the time between each quench, $t_\mathrm{exch}$ for the time between
exchange trials, $t_\mathrm{measure}$ for the time between measurement
sweeps (where the energy is sampled).

For the exchange Monte Carlo scheme to efficiently sample the
configuration space all replicas needs to tour the whole range of
temperatures in a reasonably short time. At the same time one would
not like the exchange trials, at any neighboring temperatures, to be
constantly affirmative---then the separation of the two temperatures
would be of no use. We follow Ref.~\cite{xmc} and choose the
temperature set
\begin{equation}
  T_i=T_\mathrm{low}\left(
  \frac{T_\mathrm{high}}{T_\mathrm{low}} 
  \right)^{(i-1)/(N_T-1)}~,
\end{equation}
where $1\leq i\leq N_T$ enumerates the replicas. $T_\mathrm{low}$ is
the lowest and $T_\mathrm{high}$ represent the highest temperatures
respectively. To find the actual parameter values (which will be
stated in Secs.\ \ref{sec:mod2} and \ref{sec:real_res}) one has to
check that the replicas travels throughout the temperature range with
reasonable exchange ratios for all temperature gaps.

\begin{figure}
  \centering{\resizebox*{8cm}{!}{\includegraphics{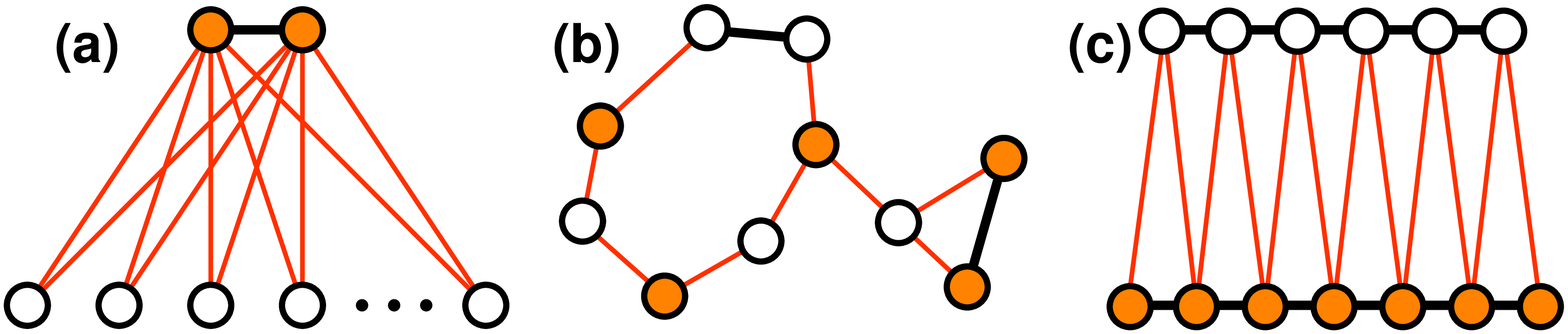}}}
  \caption{Some graphs in the discussion of the $b_2$ quantity. The
    coloring of the vertices minimizes $\mfr$. Black edges indicate
    frustration. (a) An almost bipartite graph with many
    triangles. (b) A graph where all odd-circuits contribute to the
    frustration. (c) A graph were only the shortest circuits
    contribute to the frustration.}
  \label{fig:ex}
\end{figure}

\subsection{The measure $b_2$}
Apart from finding an approximative value of $b_1$, one can also
define a quantity that is exactly solvable in polynomial time. Our
intention is in the first hand not to make a heuristic algorithm for
calculating $b_1$, but rather a quantity that captures the same
structure, i.e.\ that grows monotonously with $b_1$.

That a graph contains no odd circuits is the defining property of
bipartiteness~\cite{intro}. It is thus natural that we base a
bipartivity measure on an odd-circuit count in some
way. Unfortunately, defining a quantity in this way becomes a little
bit more complicated than at first expected. One complication is
that a graph can be very close to bipartite and still contain many
odd-circuits (see Fig.~\ref{fig:ex}(a)). A way of dealing with this
problem is to mark as few edges as possible such that each odd circuit
contains at least one marked edge. In many cases a marked edge will
correspond to a frustrated edge of the ground state of the
antiferromagnetic Ising model. In Fig.~\ref{fig:ex}(a) only the upper,
horizontal edge needs to be marked. Another problem one faces is
how to deal with odd circuits of different length---in a network with
very few odd circuits a circuit of, say, length seven would contribute
as much to the global frustration of the network as a triangle (a
subgraph of three adjacent vertices---see
Fig.~\ref{fig:ex}(b)). But in many real networks the total length of
the odd circuits is very long (this is true for all networks we
measure, see Sect.~\ref{sec:rwn}), much larger than $M$, in these cases
the short circuits are in general the most important in determining
the ground state configuration. For example, in Fig.~\ref{fig:ex}(c)
$M=23$, and while we have 11 triangles, summing the lengths of all odd
circuits gives 218 (33 from the 11 triangles, 45 from the nine circuits of
length five, and so on). However, only the triangles contributes to the
ground state configuration in the sense that each triangle has the
same configuration as the ground state of an isolated triangle, while
all odd circuits of length larger than four (e.g.\ the periphery) has not
the best coloring for a circulant of that length. To deal with this we
need to weight short circuits higher than long. We will do this by
assigning a cut-off length and neglect all circuits exceeding this
length.

\subsubsection{Definition\label{sec:def}}

Now, we make an algorithm of the above ideas as follows: Let $C_n$ be
the set of odd circuits of length $\leq n$. Let $\Sigma(C_n)$ be the
accumulated length of the circuits in $C_n$ (so, for example
$\Sigma(C_3)=3$ in Fig.~\ref{fig:ex}(b)). Now we assign the cut-off
$3M$ to $\Sigma(C_n)$, and let $\hat{n}$ be the smallest $n$
such that $\Sigma(C_n)\geq 3M$. Next we turn to the marking procedure
sketched above. Let $\nu(e)$ denote the number of circuits in
$C_{\hat{n}}$ passing through the edge $e$. Clearly edges of
high $\nu$ are likely to be frustrated in the ground state
(viz.\ Fig.~\ref{fig:ex}(a)). We now estimate $\mfr$ roughly
as the number of edges that has to be marked so that each odd circuit
of length $\leq\hat{n}$ is marked at least once. To be precise we
perform the following algorithm:
\begin{enumerate}
\item Start with $C=C_{\hat{n}}$.
\item Sort the edges in order of $\nu$.\label{step:0}
\item Repeat the following while $C\neq\varnothing$:\label{step:1}
\begin{enumerate}
\item \label{step:a} Mark the edge $e$ with highest $\nu$.
\item \label{step:b} Remove all circuits in $C$ containing $e$.
\item \label{step:c} Recalculate $\nu$ for each edge.
\end{enumerate}
\end{enumerate}
Then the number of iterations $m'$ is the assessment of $\mfr$, and we
define our bipartivity measure as
\begin{equation}
  b_2=1-\frac{m'}{M}~.
\end{equation}

This algorithm is not an attempt to actually identify the frustrated
edges, rather it is supposed to give a high $\mfr$ for a system with
high (total) geometric frustration, and vice versa: Firstly, it does
not necessarily find the minimal number of edges needed to be marked
for all odd circuits of length less than $\hat{n}$ to contain a marked
edge. But we expect this steepest descent optimization to come close
in most cases. Secondly, an odd circuit can in reality only have an odd
number of frustrated edges, but in the algorithm there is no such
restriction on the number of marked edges.

In case there are more than one edge with the highest $\nu$ (in step
\ref{step:a} of the algorithm) we choose the edge to mark at
random. The variance between different random seeds turns out to be
negligible in most cases. We will run the algorithm for different seeds to
choose the highest $b_2$ value, and get an idea about the error in
$b_2$ from the selection of edge to mark. An alternative (and more
ambitious) approach would be to iterate the whole calculation until
the highest $b_2$ has reappeared a fixed number of times (cf.\
\cite{ww}).

If we assume a sparse network (i.e.\ $N\propto M$) the running time of
the algorithm above is $O(M^2)$. To see this we first note that there
can be at most $O(M)$ iterations at step~\ref{step:1}. To find the
edge with highest $\nu$ (in step~\ref{step:a}) we do not need to sort
all edges more than once (as done in step~\ref{step:0}). Instead we
can find this out while recalculating $\nu$ (in
step~\ref{step:c}). Removing all circuits containing $e$ (as in
step~\ref{step:b}) can be done in time bounded by the total length of
circuits containing $e$, which cannot be larger than
$3M$. Step~\ref{step:c} also needs to go through all circuits passing
$e$ and thus needs the same running time as step~\ref{step:b}. To sum
this up the running time for this section of the algorithm is of order
$N^2$.

\subsubsection{Limit properties}

In the $N\rightarrow\infty$ limit the $b_2$ measure lies in almost the
same interval as $b_1$. The upper limit $b_2=1$ is attained if and
only if the graph is bipartite. (If the graph is bipartite
$C_{\hat{n}}$ is empty and $\nu(a)=0$ for all $a$, so $m'=0$ and
$b_2=1$. If there exists odd circuits $m'\geq 0$, so $b_2<1$.) $b_2$
cannot be as low as 0 (if one marks all edges, every circuit must be
marked). Since the $b_2$-definition is inspired by the ground-state
configuration of the antiferromagnetic Ising model, we expect a
similar lower bound to $b_2$ as to $b_1$. In Appendix~\ref{sec:bound}
we argue that the lower bound on the $b_2$, as for the $b_1$ measure,
is $1/2$ in the $N \rightarrow \infty$ limit.

\subsubsection{The complete algorithm}
So far we have overlooked the central part in calculating the $b_2$
measure---namely to find odd circuits. To do this we use a modified
version of Johnson's algorithm~\cite{johnson}. In principle Johnson's
algorithm is a depth first search where, to avoid futile searching,
some vertices are blocked while stepping down the search tree. The
running time for Johnson's algorithm is $O(M(C+1))$ (if $M>N$) where
$C$ is the total number of circuits. Now $C$ can grow fast with
$N$ which would make the finding of all odd circuits a quite
intractable computation. In many cases the cut-off of the circuit length,
that we introduced above to give less priority to long circuits, saves
us by setting a limit on the search depth. To implement this we
let $\bar{n}$ be the current upper bound on circuit length (or search
depth), and $\bar{\Sigma}$ be the current sum of odd circuits
$\leq\bar{n}$. As soon as $\bar{\Sigma}\geq M$ we iteratively
decrease $\bar{n}$ by $2$ and recalculate $\bar{\Sigma}$ until
$\bar{\Sigma}<3$. If $\bar{\Sigma}< M$ when the search is over we
rerun the procedure where we use $\bar{n}+2$ as our new (fixed)
$\bar{n}$~\cite{note:alt}. When the search is over we assign
$\hat{n}$ the value $\bar{n}$. For dense bipartite graphs the
algorithm is intractable. In the worst case, the full bipartite
graph, $K_{N/2,N/2}$, there are
\begin{equation}
  C(K_{N/2,N/2})= \sum_{k=4}^{N}\frac{1}{2k}
  \left[\frac{(N/2)!}{(N/2-k/2)!}\right]^2
\end{equation}
circuits (where the sum is over even values of $k$)~\cite{note:bip}
giving a running time of $O(N^2C(K_{N/2,N/2}))$. One can of course
decide whether or not a graph is bipartite in linear time, but
non-bipartite cases of similar complexity are easily constructed (by,
e.g., adding an isolated triangle). In practice these worst cases are,
probably, very rare---a, relatively speaking, very low density of odd
circuits is needed to get a small $\hat{n}$---even in the real-world
network with highest bipartivity we have $\hat{n}=3$. In this case
($\hat{n}=3$) all odd circuits are found in $O(M^2)$ time.

Now we turn to a more complete description of the algorithm. Johnson's
algorithm takes the `least' (smallest in some enumeration) vertex in a
strongly connected subgraph as its starting point. To find strongly
connected components we use the algorithm in Ref.~\cite{SCC}. To sum
up, the algorithm reads:
\begin{enumerate}
\item Mark all vertices as unchecked.
\item While there are unchecked vertices, iterate the
  following:\label{step:wh}
\begin{enumerate}
\item Pick an unchecked vertex $v$.
\item Find the largest strongly connected component $\Lambda_v$
  containing $v$.
\item Set $\Lambda:=\Lambda_v$ and repeat the following steps as long
  as $\Lambda\neq\varnothing$:
\begin{enumerate}
\item Pick the least vertex $u$ of $\Lambda$.
\item Call a subroutine implementing the modified Johnson's
  algorithm. Recalculate $\bar{n}$ and add $C_{\bar{n}}$ to a list
  $\mathcal{C}$. Delete circuits longer than $\bar{n}$ from $\mathcal{C}$.
\item Delete $u$ from $\Lambda$.
\end{enumerate}
\end{enumerate}
\item Set $\hat{n}:=\bar{n}$.
\item Run the algorithm described above (in Sect.~\ref{sec:def}) to
  mark edges and calculate $b_2$.\label{step:calc}
\end{enumerate}
In all cases, step~\ref{step:wh} sets the limit on running time. As
mentioned, in most application we expect the running time of
step~\ref{step:wh} to be $O(M^2)$ (similarly to that of
step~\ref{step:calc}).

\section{The Networks}

\begin{figure}
  \centering{\resizebox*{8cm}{!}{\includegraphics{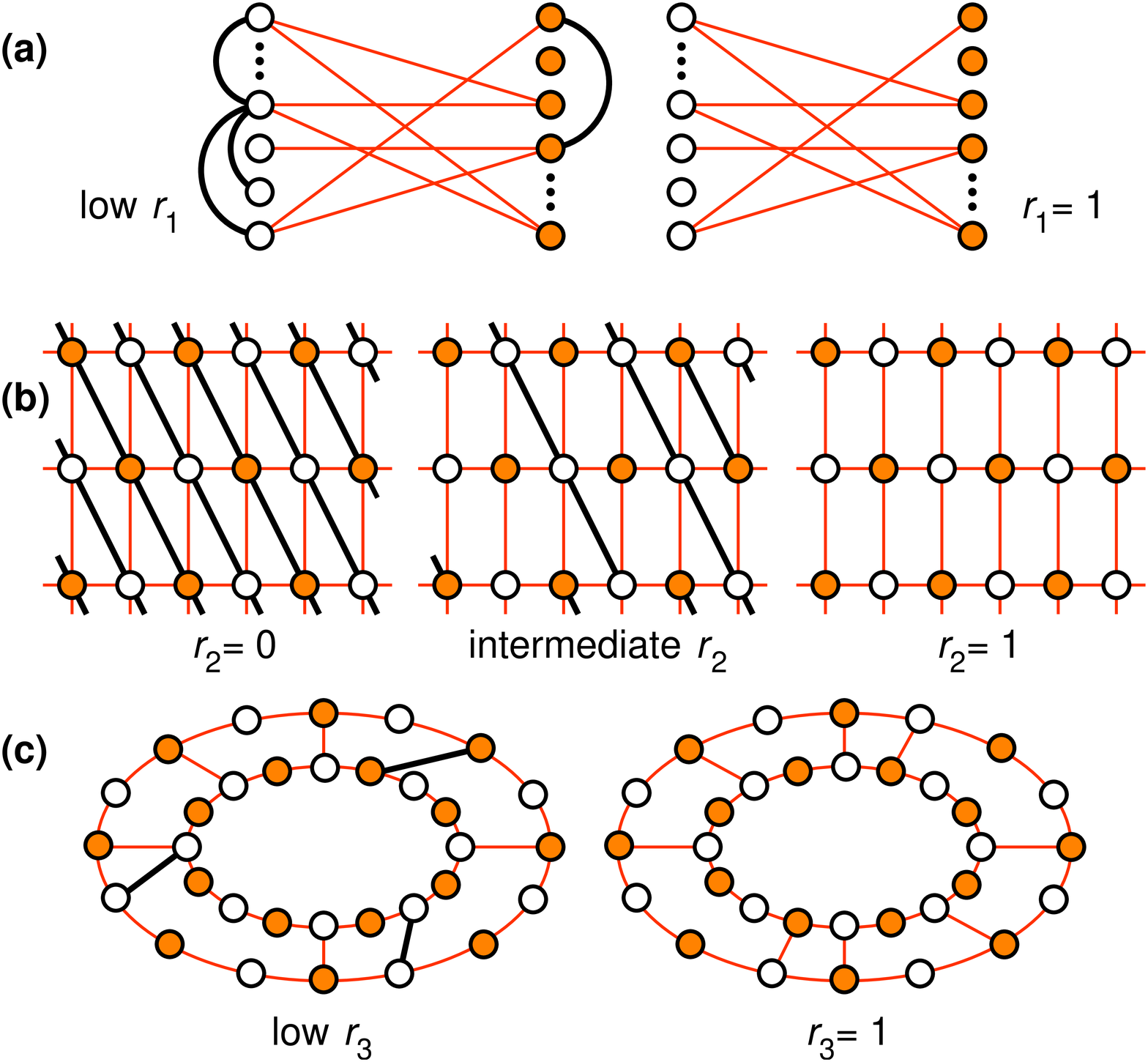}}}
  \caption{Construction of the test networks. (a) shows the
    generalization of the ER model (Model 1). (b) shows interpolation
    between quadratic and triangular lattices (Model 2). (c) shows the
    model with predominantly longer circuits (Model 3). All models
    are bipartite for $r_{1,2,3}=1$. Additional edges creates
    odd circuits (frustration) for lower $r_{1,2,3}$-values. The black
    lines illustrates these additional edges. The white and non-white
    vertices symbolize a partition giving $b_1=1$ in the $r_{1,2,3}=1$
    case (it is not meant to represent the optimal coloring when
    $r_{1,2,3}<1$).
  }
  \label{fig:mo}
\end{figure}

\subsection{Test networks with tunable bipartivity\label{sec:mod}}
To test and compare the $b_1$ and $b_2$ quantities we construct three
types of test networks where the bipartivity can be tuned by model
parameters. The principle behind all models is to start from
bipartite networks and add lesser or greater number of edges within a
partition to create odd circuits.

One type (Model 1) is a quite straightforward generalization  of the
Erd\"{o}s-Renyi (ER) model~\cite{ER}: We partition the vertices in two
disjoint sets of sizes $\tilde{N}$ and $N-\tilde{N}$. Then we add $r_1
M$ edges randomly between vertices of the different sets, and
$(1-r_1)M$ edges regardless of what set the vertices belongs to (see
Fig.~\ref{fig:mo}(a)). In this way we interpret $r_1$ as the
strength of the heterophilous preference in a model where bipartivity
is the only structural bias. The choice of vertex pairs
is done with randomness, the only restriction being that loops and
multiple edges are not allowed. If $r_1=0$ the model reduces to the ER
model, while for $r_1=1$ the networks are bipartite (cf.\
Ref.~\cite{nws}).  This model is probably the most random (i.e.\
having least structural biases) model with tunable bipartivity. The
disadvantage is that the expectation values of $b_1$ and $b_2$ are
hard to calculate (even in the frustrated limit $r_1=0$).

Model 2 interpolates between two-dimensional square- and triangular
lattices. We start, for $r_2=0$, with a triangular grid with periodic
boundary condition. Let $L$, the linear dimension of the system (i.e.\
$N=L^2$), be even. For a non-zero parameter value we (by uniform
randomness) delete $r_1L^2$ `diagonal' edges creating frustration as
illustrated in Fig.~\ref{fig:mo}(b). To be more precise, if we index
the vertices as $(i_x,i_y)$, $1\leq i_x,i_y\leq L$; then the edges are
$[(i_x,i_y),(i_x+1,i_y)]$ and $[(i_x,i_y),(i_x,i_y+1)]$ (giving the
square grid) plus $r_1L^2$ edges of the form
$[(i_x,i_y+1),(i_x+1,i_y)]$ chosen by uniform randomness (addition is
modulo $L$). This model has a high degree of short circuits. The
extremes $r_2=0$ and $r_1=1$ represent two generic lattice types. The
symmetries of the regular networks simplify the calculations of
e.g.\ limit properties for the bipartivity measures. If $r_2=1$ the
system is bipartite (note that $L$ has to be even for this to hold) so
$b_{1,2}=1$. When $r_2=0$ we have $b_1=b_2=2/3$:  For the lower limit
of the $b_1$ quantity, see Ref.~\cite{WAHO}. For the lower limit $b_2$
we note that $\Sigma(C_3)=6N$ (since each vertex can be associated
with two triangles). This gives $\hat{n}=3$ and $\nu=2$ for all
edges. Now it is enough to mark $N$ edges (e.g.\ all
$[(i_x,i_y),(i_x+1,i_y)]$ edges). In this case we note that each edge
will have $\nu=2$ when it is marked, which means that the marking
sequence is optimal and that the number of iteration cannot be less
with another choice of edges to mark. So $b_2=1-N/3N=2/3$. The major
disadvantage with Model 2 is that the average degree is a function of
$r_2$ ($M=(3-r_2)L^2$). This change in the average degree can make it
harder to separate effects of the shift in bipartivity from the shift
in average degree.

In both model 1 and (even more) model 2 triangles will dominate
the set of odd circuits. To test networks with predominantly longer
circuits we construct a Model 3 as follows (see Fig.~\ref{fig:mo}(c)):
We make two circulants of size $N/2$ with the vertices
$\{v_1^i,\cdots,v_{N/2}^i\}$ and edges
$\{(v_1^i,v_2^i),\cdots,(v_{N/2-1}^i,v_{N/2}^i), (v_{N/2}^i,v_1^i)\}$,
$i\in\{1,2\}$. Then we add $M_\mathrm{trans}$ transverse edges between
the circulants. $M_\mathrm{trans}/2$ of these edges are placed out
separated by equal distance $N/M_\mathrm{trans}$ separating the double
circulants into $M_\mathrm{trans}/2$ `sectors.' Then we fill up each
sector with another transverse edge: With probability $r_3$ we add an
$(v^1_i,v^2_i)$ edge (such that $(v^1_i,v^2_i)$ is none of the
previously added transverse edges), otherwise we add a
$(v^1_i,v^2_i+1)$ edge (addition modulo $N/2$). We note, to a first
approximation, that if $r_3=0$ marking (in the process of calculating
$b_2$) one edge between every transverse edge on one of the circulants
is needed to mark the shortest odd circuits. This will make $b_2\in
O(1-M_\mathrm{trans}/N)$.

\subsection{Real-world networks\label{sec:rwn}}
Physicists' networks studies has, in the spirit of statistical mechanics,
emphasized the properties remaining when the system grows beyond any
limit. Bipartivity, as discussed above, is well defined for all system
sizes. Still it is a quantity that can potentially suffer from
finite-size effects (from the fact that not all real neighbors of all
actors in a empirically constructed social network are a part of the
graph) and is therefore preferably measured for large networks. Now
the problem is to find data for large-scale real-world networks of
social interaction. In general two methods has been successful for this
purpose---one either uses professional collaborations of some sort or
data from interaction over the Internet (either in Internet
communities~\cite{HEL,smith}, or through email exchange~\cite{ebel}.

\subsubsection{Professional collaboration networks}
In the professional collaboration networks we study the
vertices are professionals of some field---networks of scientists
and company directors are considered in this papers, the movie-actor
network is another frequently studied example; the edges represent
that two actors has been involved in the same professional
collaboration. This is some-times referred to as a ``one-mode''
representation of an affiliation network (as opposed to the bipartite
two-mode representation discussed in Sect.~\ref{sec:intro}).

Professional collaboration networks are no doubt interesting in their
own right as accounts for the interaction dynamics of the respective
fields. Assuming that the formation of professional ties follow
similar principles as general human interaction, we can use
professional collaboration networks to draw conclusions about the
structure of more general social networks. However, at one point (at
least) professional collaboration differs from general social
interactions: A collaboration tie does not necessarily imply a strong
personal acquaintance, but in these networks each collaboration
constitutes a fully connected cluster. This leads to higher fraction
of short circuits than, say, a friendship network.

One of the professional collaboration network we use is of
scientists who has uploaded manuscripts to the preprint repository
arxiv.org. Two scientists are linked if their name (identified by
surname and initials) appear together on at least one preprint. A
detailed description of this network can be found in
Ref.~\cite{newman3}. In the other professional collaboration network
the vertices represent company directors from the Fortune top 1000
list of companies in USA the year 2001. An edge (collaboration) in
this network means that two directors are sitting in board of the same
company. A detailed description of this network can be found in
Ref~\cite{davis}. Sizes of the networks can be seen in
Table~\ref{tab:b}.

\subsubsection{Online interaction networks}

In online interaction networks, the vertices are users of Internet
communities and an arc (A,B) is added if A contacts B, or
if A adds B to his/her list of friends~\cite{smith,HEL}. Another kind
of online interaction networks are email networks~\cite{ebel}, where
an arc can be assigned if an email is sent, or if a person adds
another to his/her address book. Just as for professional collaboration
networks, one can argue that online interaction networks are
representative as general social networks. One can assume that new
contacts are formed through preference-matching searches to a larger
extent, and introduction by mutual friends to a lesser extent, than in
general friendship networks. Since the introduction of mutual friends
to each other is believed to be the major cause of high clustering
(large density of triangles, or, large transitivity)~\cite{newman1}
one can expect a lower clustering in networks of online interaction
(still the clustering in these network seems to be finite in the
$N\rightarrow\infty$ limit~\cite{HEL}).

The specific online interaction networks we consider are constructed
from the Internet communities nioki.com and pussokram.com. The
nioki.com data is described in Ref.~\cite{smith}. In this data an arc
(A,B) means that B is listed as a friend by A, which
allows A to see if B is online and send instant messages to B. In
the pussokram.com data the arcs correspond to communication between
the users. There are four different types of communication in this
specific network (all described in detail in Ref.~\cite{HEL}). We use
the networks obtained from two types of interaction (`messages'---like
ordinary emails within the community, and ``guest book''---where one user
contacts another by writing in his/her guest book), and the network
of any of the four types. Network sizes can be found at
Table~\ref{tab:b}.

Another large difference between the pussokram.com and nioki.com data
is that the former community has a very pronounced romantic profile,
encouraging flirts and romantic correspondence. nioki.com has also a
search engine to ``trouve l'amour'' (find love), but that is all.

Apart from the two Internet communities, we study another type of
online interaction network based on the flow email. For this network
all in- and out-going email traffic to a server was logged for around
three months~\cite{ebel}. The server handles undergraduate students'
email accounts at Kiel University, Germany. Thus there are two
categories of vertices---internal vertices, whose activity is accurately
mapped; and external vertices, that only have edges leading to internal
vertices. In this study we restrict ourselves to the network of
internal-internal contacts. The reason we do not include external
contacts is that we would miss the (probably many) circuits containing
external-external edges which would bias the bipartivity.

\subsubsection{Network from interview and field survey\label{sec:soc}}
Apart from the above networks, all obtained from databases, we also
measure the bipartivity of two networks obtained from interview and
field surveys. The first data set is gathered by observations of
interaction between members of a university karate
club~\cite{zach}. We also study the network of acquaintance ties in a
prison~\cite{prison}. The outgoing arcs from A corresponds to
prisoners listed by A in response to the question: ``What fellows on
the tier are you closest friends with?'' Due to their acquisition
methods these kind of real-world networks has to be rather small. This
can, as mentioned, result in finite size effects. On the other hand
they, most likely, more truly reflect the structure of real
acquaintance networks.

\section{Results}

In this section we present the results of the test networks and the
measurement for the real-world social networks.

\subsection{Test networks\label{sec:mod2}}

\begin{figure}
  \centering{\resizebox*{8cm}{!}{\includegraphics{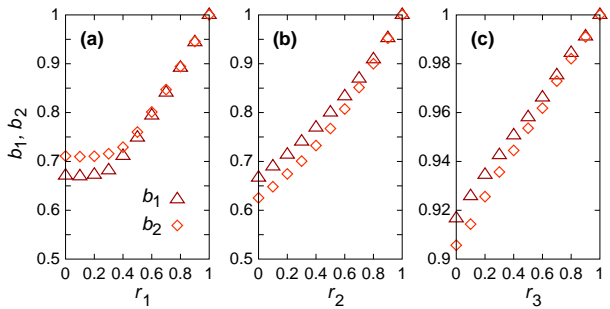}}}
  \caption{The bipartivity measures versus the model parameters of the
  two models defined in Section~\protect\ref{sec:mod}. (a) shows the
  result for Model 1, (b) shows the result for Model 2, and (c) shows
  the result for Model 3. All error bars would be smaller than the
  symbol size. The monotonous growth of the bipartivity measures shows
  that the measures behaves expectedly.}
  \label{fig:mx}
\end{figure}

As expected, both $b_1$ and $b_2$ are monotonously increasing as
functions of the $r_1$, $r_2$ and $r_3$ parameters of
(almost~\cite{note:not_really} all our test network (see
  Fig.~\ref{fig:mx})). This is encouraging and suggests that both
$b_1$ and $b_2$ are quite relevant measures of bipartivity.

The Model 1 measurements shown in Fig.~\ref{fig:mx}(a) are made with the
model parameters $N=2\tilde{N}=100$ and $M=800$. We have checked many
other sizes too, but all have the characteristic appearance of
Fig.~\ref{fig:mx}(a)---a linear increase of $b_1$ and $b_2$ for larger
$r_1$ and an flatter slope for $r_1$ close to zero. This shape is
expected from the discussion in Sect.~\ref{sec:intro}---in networks
where a heterophilous preference is the only structure-inducing
force, only the strong preference limit gives a strong measurable
effect: Close to the ER limit $r_1\approx 0$, the original two
partitions will not be identified correctly, only when the different
partition (to a large extent) have different sign the bipartivity will
be proportional to the strength of the heterophilous preference.

As seen in Fig.~\ref{fig:mx}(b) Model 2 shows an almost linear
functional form of $b_{1,2}(r_2)$. In this case, triangles dominate
the odd circuits even at small values of $r_2$. Tuning $r_2$ will give
a proportional increase of the number of triangles. Thus a linear
$r_2$ dependence of $b_2$ would be expected.

Also Model 3 has linear $b_{1,2}$ vs.\ $r_3$ curves. The model
parameters used are $N=100$ and $M_\mathrm{trans}=10$. As mentioned in
Section~\ref{sec:mod}, we expect $b_2\approx M_\mathrm{trans}/N$ for
$r_3=0$, which is confirmed in Fig.~\ref{fig:mx}(c).

The measurements for both $b_1$ and $b_2$ are averaged over 100
network realizations. The XMC scheme for the $b_1$ quantity is ran at
24 temperatures in parallel, between temperatures $0.01$ and
$2$. Other network parameters are $t_\mathrm{avg}=4\times 10^5$,
$t_\mathrm{measure}=4$, $t_\mathrm{quench}=20$ and
$t_\mathrm{exch}=1000$. These are more modest parameter values than we
will use for the real-world networks, but the test networks are also
much smaller, and since the distribution of $b_1$ and $b_2$ are
(likely) symmetric, the network average helps to reduce the error.

\begin{table*}
  \caption{Sizes, clustering coefficients and bipartivity measures
  $b_1$ and $b_2$ for real-world social networks.}
\label{tab:b}
\begin{ruledtabular}
\begin{tabular}{l|rrr|dddd|dddd}
  network & $N$ & $M_\mathrm{dir}$ & $M$ & C_\mathrm{dir} & C &
  D_\mathrm{dir} & D & b_1^\mathrm{dir} & b_1 &
  b_2^\mathrm{dir} & b_2 \\\hline 
  all contacts & $29{\:}341$ & $174{\:}662$ & $115{\:}684$& 0.012 &
  0.0060 & 0.016 & 0.017 & 0.859 & 0.860 & 0.948 & 0.928\\
  messages & $20{\:}691$ & $73{\:}346$ & $52{\:}435$& 0.0052& 0.0061 &
  0.0081 & 0.0061  & 0.897 & 0.892 & 0.984 & 0.964\\
  guestbook & $21{\:}545$ & $76{\:}257$ & $55{\:}076$& 0.014 & 0.014 &
  0.015 & 0.021  & 0.863 & 0.889 & 0.943 & 0.965\\
  nioki.com & $50{\:}259$ & $405{\:}742$ & $239{\:}452$& 0.0076 &
  0.0065 & 0.016 & 0.013  & 0.842 & 0.855 & 0.956 & 0.975\\
  emails & 637 & 554 & 443& 0.11 & 0.16 & 0.071 & 0.14 & 0.944 & 0.944
  & 0.971 & 0.941 \\ 
  arxiv.org & $52{\:}909$ & $\times$ & 490{\:}600 & \times & 0.45 &
  \times & 0.35 & \times & 0.630 & \times & 0.623\\
  directors & 7${\:}475$ & $\times$ & 48{\:}899 & \times & 0.21 &
  \times & 0.37 & \times & 0.549 & \times & 0.507\\
  karate club & 34 & $\times$ & 78& \times & 0.26 & \times & 0.26  &
  \times & 0.782 & \times & 0.782 \\ 
  prison & 64 & 182 & 85& 0.19 & 0.31 & 0.089 & 0.14 & 0.786 & 0.878 &
  0.918 & 0.847 
\end{tabular}
\end{ruledtabular}
\end{table*}

\subsection{Real-world social networks\label{sec:real_res}}

Now we turn to the result for the bipartivity measures of real-world
networks. The values are presented in Table~\ref{tab:b}. For
comparison we also give values for the clustering coefficient (density
of triangles) $C$ and the density of squares $D$ in both directed and
undirected versions~\cite{note:cd}. Undirected networks are constructed
by taking the reflexive closure. At first glance at the table
we arrive at the pleasing conclusion that the bipartivity for the
pussokram.com networks is very high (as expected from a network of
romantic interaction of mostly heterosexuals). But disappointingly,
the bipartivity measures show similarly high values for the nioki.com
and email networks. This can be explained by the fact that nioki.com,
just like the pussokram.com, data has very low $C$ and $D$ values, and
presumably very few circuits at all. Now branches (subgraphs without
circuits that can be isolated by cutting one edge) does not give a
positive contribution to either $b_1$ or $b_2$, no matter of the
gender of the agents. The email network do have a high clustering, but
still rather high bipartivity. The reason is that the email network is
rather heavily fragmented and contains many isolated subnetworks of
two vertices and one edge, and three vertices and two edges. Such
subnetworks does not affect the clustering coefficient but tends to
decrease the bipartivity measures~\cite{note:improvement}.

The collaboration networks consist of a number of fully connected
clusters (corresponding to a specific collaboration) that are
interconnected. It is thus natural that we see low bipartivity and a
high density of short circuits. The lower bipartivity values for the
company director network can be explained by smaller average size of such
fully-connected clusters: The average number of vertices per
collaboration is $9.5$ for the corporate director network and $2.5$
for the scientific collaboration data~\cite{davis,newman3}.

The two small networks constructed from field surveys (the ``karate
club'' and ``prison'' network of Table~\ref{tab:b}, discussed in
Section~\ref{sec:soc}) show mid-range bipartivities and relative high
values of $C$ and $D$. From the above discussion we can expect that
the bipartivity of large, real, acquaintance networks is somewhere
between those of the collaboration networks and the Internet community
networks (because they probably have higher clustering than Internet
community networks, and lower number of fully connected clusters than
the collaboration networks). Encouraging enough, this is exactly what
we see in Table~\ref{tab:b}. Of course, the very small systems sizes
might affect the results, but that the bipartivity measures of
real-world acquaintance measures would be close to either the upper or
lower limits seems hard to believe.

We conclude this section by a note on the parameters for the XMC
optimization. The measurement of $b_1$ for all real-world network
(except the nioki.com data where we study the convergence more
carefully) are done just once with the following simulation
parameters, $N_T=24$ (with temperatures from $0.002$ to $5$)
$t_\mathrm{avg}=1\times 10^7$, $t_\mathrm{measure}=16$,
$t_\mathrm{quench}=40$ and $t_\mathrm{exch} = 2\times 10^4$.

\section{Summary and discussion}

This paper concerns the quantification of the network structure
`bipartivity'---how close to bipartite a given graph is. We propose
two measures for this quantity. One quantity $b_1$ based on the
optimal two-coloring of the network---or, equivalently, the ground
state of the antiferromagnetic Ising model on the network. The
exact value of this quantity (that has been used in different roles
elsewhere) is NP-complete and thus in general not feasible
to calculate exactly. Instead we seek an approximate solution by a
simulated annealing approach. The simulated annealing is based on the
exchange Monte Carlo scheme. We argue that this unorthodox
minimization method helps us avoid local minima of the energy
landscape of the antiferromagnetic Ising model. Furthermore we develop
a measure $b_2$ based on the count of odd circuits that, for almost
all networks, is calculable in polynomial time.

We propose three different random graph test models where one can
interpolate between arguably non-bipartite and bipartite graphs by
tuning a control parameter. Both our bipartivity measures are shown to
increase monotonically with tuning the control parameters towards the
bipartite extreme. From this we conclude that the bipartivity measures
really quantify the notion of bipartivity.

By considering example networks we infer that bipartivity is a
structure that cannot be measured by currently popular structural
measures, such as the clustering coefficient. At the same time any
sensible quantification of bipartivity probably has to have a positive
correlation with the clustering coefficient for most networks (with
exceptions for exotic cases like Fig.~\ref{fig:ex}(a))---so, in that
case bipartivity and clustering is not independent.

We measure $b_1$ and $b_2$ of a number of real-world networks,
constructed from online interaction, professional collaborations, and
field surveys. As expected, we see high bipartivity values for data
from the Internet community pussokram.com, where romantic contacts
are encouraged, and hence a high degree of heterophilous interaction
expected. We also see the expected low bipartivity values for the
professional collaboration and empirical acquaintance networks we
study. Disappointingly we cannot use our bipartivity measures to
distinguish between the networks driven by romantic or friendship (or
professional) contacts. To do this other structures and the network
sizes has to be taken into account, in a more elaborate analysis (that
is out of the scope of this study).

So far our examples of networks with high bipartivity has been
romantic networks and networks of sexual contacts. Network-based
studies of sexually transmitted diseases~\cite{lea} is a potentially
interesting area for bipartivity measures, as the transmission rates
for homosexual and heterosexual contacts differ~\cite{anma}. Apart
from romantic and sexual networks, there are other areas where the
bipartivity measure may prove useful: One can consider a trade network
where some agents are more or less pronounced sellers and others are
primarily buyers (cf.\ Ref.~\cite{white}), such networks would not
have a neutral bipartivity. Another application is for the
`genealogical' network of a disease outbreak: Some contagious diseases
have a relatively stable duration between when an individual is
infected and when he or she becomes infectious. Epidemics of these
types of diseases can therefore roughly be divided into different
generations of infected individuals~\cite{anma}. A network
consisting of possible edges of infections, for an outbreak of this
type of disease, should therefore have very few odd-length
circuits. The reason is that the infection is only transmitted between
succeeding generations, which generates only circuits of even length
(in the reflexive closure of the network). When reconstructing the
paths this kind of disease has taken in a population, a minimization
of the bipartivity measures can be a method for excluding redundant
infectious edges.

We conclude by an analogy to linear algebra---we have identified a new
dimension (structure) and proposed base vectors (measures), that
unfortunately are not orthogonal to the other dimensions.

\section*{Acknowledgements}
We would like to thank Niklas Angemyr, Stefan Bornholdt, Gerald Davis,
Holger Ebel, Michael Lokner, Stefan Praszalowicz, and Christian
Wollter for help with data acquisition; and Johan Giesecke, James
Moody, Mats Nyl\'{e}n, and Pontus Svenson, for comments and
suggestions. P.H.\ was partly supported by the Swedish Research
Council through contract no.\ 2002-4135. F.L.\ was supported
by the National Institute of Public Health. C.R.E.\ was supported by
the Bank of Sweden Tercentenary Foundation. B.J.K.\ was supported by
the Korea Science and Engineering Foundation through Grant No.\
R14-2002-062-01000-0.

\appendix

\begin{figure}
  \centering{\resizebox*{8cm}{!}{\includegraphics{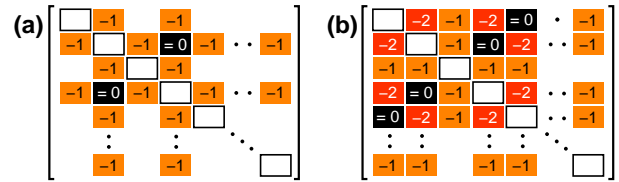}}}
  \caption{Marking of edges (in matrix representation) while
    calculating the $b_2$ quantity for a fully connected graph. `$-1$'
  means that $\nu$ at that position is decreased by one unit, `$=0$'
  means that $\nu=0$ at that position.
  }
  \label{fig:ma}
\end{figure}

\section{The lower bound of the measure $b_2$\label{sec:bound}}

In this Appendix we argue that, in the $N\rightarrow \infty$ limit,
the lower bound for $b_2$ is $1/2$ (just like $b_1$). First we
conjecture that the minimal value for $b_2$, just as for $b_1$, is
attained for complete graphs. (This will be further motivated below.)

To assess $b_2$ for complete graphs, we note that~\cite{note:sigsum}
\begin{subequations}
\begin{eqnarray}
  \Sigma(C_n)&=&\sum_{\mathrm{odd}\;3\leq i\leq n}
  \frac{N!}{2(N-i)!}~\Rightarrow\label{eq:sigsum}\\
  \Sigma(C_3)&=&\frac{N(N-1)(N-2)}{2}\geq\nonumber\\&\geq&
  \frac{N(N-1)}{2}=M~,
\end{eqnarray}
\end{subequations}
so $\hat{n}=3$ which results in that $\nu=N-2$ for each edge.

Now we apply the marking procedure of Sec.~\ref{sec:def}. Marking an
edge $(u,v)$ makes $\nu(u,v)= \nu(v,u)=0$. Furthermore, every edge
$(u,w)$ and $(v,w)$ ($w\neq u,v$) will be decreased by one since the
triangle $\{u,v,w\}$ now contains a marked edge. The discussion will
be simplified by considering a matrix representation of $\nu
(u,v)$. Marking $(u,v)$ sets $\nu(u,v)=\nu(v,u)=0$ and decreases the
$u$'th and $v$'th columns, and $u$'th and $v$'th rows by one (an
example is given in Fig~\ref{fig:ma}(a)). Marking another edge
$(u',v')$ ($u'$ and $v'$ are different from both $u$ and $v$,
otherwise $\nu(u',v')$ would not be maximal) will have the same effect
as marking the first. For positions like $(u,v')$ the original $\nu$
are decreased by 2 (see Fig.~\ref{fig:ma}(b)), since it has lost the
two passing triangles $\{u,u',v'\}$. and $\{v',u,v\}$. Continuing this
process we see that it takes $N/2+O(1)$ markings for $\nu$ of each
edge to be decreased by two units, and thus $m'=N^2/4+O(N)$ markings to
make $\nu=0$ for all edges. This gives $b_2=1/2$ in the $N\rightarrow
\infty$ limit. Since the appropriateness of $b_2$ as a bipartivity
measure is not really dependent on the limit values, we will not give
a rigorous proof that the correction is of a lower order for all
levels of the marking procedure (one level is the $N/2+O(1)$ edges
needed to be marked for $\nu$ to be decreased by at least two units
for each edge).

\begin{figure}
  \centering{\resizebox*{7.5cm}{!}{\includegraphics{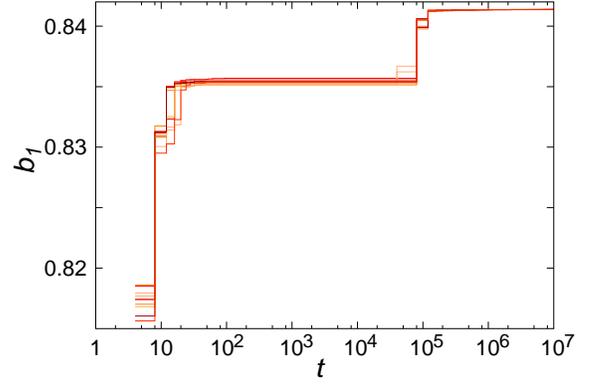}}}
  \caption{The current value of $b_1$ (at the lowest-temperature level
    of the cooling) as a function of running time for ten independent
    measurements of the directed version of the nioki.com data.
  }
  \label{fig:co}
\end{figure}

Now we argue that the $b_2$ takes its minimal value for complete
graphs. First we note that the number of circuits of length $n$ per
edge, for any $n$, is largest in a complete graph~\cite{review2}. So
if we set $\hat{n}$ arbitrarily and discard circuits of length $\leq
n$ in the calculation of $\nu(v)$, the fully connected graph would
give the highest $m'$ value and thus the lowest bipartivity
measure. The strongest candidate for a lower bipartivity measure than
that of a fully connected graph would thus be a graph such that the
$\Sigma(C_n)< 3M$ and $\Sigma(C_{n+2})$ is as big as possible for some
$n$. But the number edges needed to be removed from a fully connected
graph for $\Sigma(C_n)< 3M$ to hold, not only reduces the contribution
to $\nu$ from circuits of length $n$ but also from circuits of length
$n+2$ to a similar extent. If one performs the approximate marking
procedure outlined above for circuits of length five one starts from
$\nu=(N-2)(N-3)(N-4)$ and it takes $N/2+O(1)$ markings to decrease
every $\nu$ with at least $2N^2$. This means that the number of edges
needed to be marked to make $\nu = 0$ for every edge is the same if
circuits of length five is considered. It also means that a graph as
outlined above (with $\Sigma(C_n)< 3M$ and $\Sigma(C_{n+2})$ is as big
as possible) probably do not have a lower $b_2$ than a complete graph.

To epitomize, the $b_2$ measure lies in the interval $[1/2,1]$ in the
$N\rightarrow \infty$ limit. The finite size corrections to $b_2$ for
fully connected graphs, however, turns out to make $b_2$ slightly less
than $1/2$.

\section{Convergence of the simulated annealing\label{sec:simann}}

To analyze the convergence of the simulated annealing scheme we run
ten independent calculations of the $b_1$ quantity (with the same
parameter values as in Sect.~\ref{sec:real_res}). The individual time
evolutions of $b_1$ (at the lowest temperature $T=0.002$) for the
different runs are shown in Fig.~\ref{fig:co}. We note that already
after the first quench $b_1$ is only $3\%$ away from the value at the
end of the run, and after 50 time steps $b_1$ is $0.5\%$ of the value
after $1\times 10^7$ time steps. We note that there is no way of
constructing a statistically valid confidence interval for the true
$b_1$ value since an arbitrary complex energy landscape could have a
global minimum with a basin of attraction of measure zero. There are
however indications that this is seldom a major problem, at least not
for the bisection problem~\cite{jerrum}.

An interesting observation from Fig.~\ref{fig:co} is the step-like
structure. This is a result of the exchange trials: After $t\approx
100$ the local minimum has been found, but at the temperature in
question the system is in principle stuck in a confined part of the
configuration space, and cannot enter lower lying energy
valleys. In the time scale $t = 10^5$ there is another jump in the
$b_1$ value. This is related to that other replicas from other parts
of the configuration space reaches the lowest level. At around
$t=10^6$ the current highest $b_1$ values (lowest energy) reaches
another plateau. At this time, each replica should have covered the
whole temperature range several times. This second plateau gives two
encouraging implications: Firstly, that the correct value of $b_1$
probably is not very far off the measured value. Secondly, that the
exchange steps really are helpful. If one wants to run this algorithm
more efficiently the $t_\mathrm{exch}$ we use is far too large (but
beneficial for separating the time scales in the discussion
above). Ideally $t_\mathrm{exch}$ should probably be chosen to be of
the same order as the first jump (from the regular Monte Carlo
steps)---in the nioki.com network (displayed in Fig.~\ref{fig:co})
this would be $t\approx 100$.


\begin{thebibliography}{99}
\bibitem{note:alt} The intuitive way to find a least upper bound might
  be to search at two levels simultaneously ($\bar{n}$ and
  $\bar{n}-2$) and decrease the bound $\bar{n}\mapsto\bar{n} -2$ when
  $\Sigma_{\bar{n} -2}\geq M$ ($\Sigma_{\bar{n} -2}$ denotes the sum
  of the length of all circuits shorter than or equal to $\bar{n}
  -2$. This would slow down the computation considerably since our
  modified Johnson's algorithm mostly finds circuits of the length of
  the search depth $\bar{n}$, and thus it takes a long time to
  increase $\Sigma_{\bar{n} -2}$.
\bibitem{note:bip} Consider $K_{N/2,N/2}=(V,U,E)$ where $V$ and $U$
  are the two vertex sets. We write a circuit as a $k$-tuple
  $(v_1,u_1,\cdots,v_{k/2},u_{k/2})$ where $v\in V$ and $u\in U$. Then
  there are $(N/2)\,(N/2)\,\cdots\, [N/2-(k/2-1)]\,[N/2-(k/2-1)] =
  [(N/2)!/ (N/2-k/2)!]^2$ distinct $k$-tuples. As for circuits, the
  choice of start-vertex $v_1$ does not matter, neither does the
  direction matter. To compensate for this we divide by $1/2k$ to get
  the right number of circuits of length $k$ in $K_{N/2,N/2}$.
\bibitem{note:not_really} Actually the $b_1$ value for Model 1 is
  $0.2\%$ lower (around three standard deviations) for $r_1=0.1$ than
  for $r_1=0$. We will not speculate in the reason for this since the
  effect is small and the overall picture is clear.
\bibitem{note:cd} If one denotes the number of representations of
  (directed or undirected) circuits of length $n$ by $c(n)$ and the
  number of representations of paths of length $n$ by $p(n)$. (By
  representations we mean different ways of listing adjacent vertices;
  so, for example, a triangle has six representations.) Then we can
  can define $C=c(3)/p(3)$ and $D=c(4)/p(4)$. For more detailed
  definitions, see Refs.~\cite{bw} and \cite{HEL}.
\bibitem{note:improvement} A potential improvement would be to measure
  $b_1$ and $b_2$ on the 2-core (the maximal subgraph with minimal
  degree 2) of $G$. This would eliminate circuit-free subgraphs that
  contains no information about the degree of heterophilous preference
  among the agents forming the network.
\bibitem{note:sigsum} In the $K_N$, the number of circuits of length
  $i$ is the $i$-permutations $N!/(N-i)!$ divided by $2i$ (a factor
  $i$ to compensate for the over-counting since a circuit is
  independent of starting vertex; a factor 2 to compensate for the
  double counting of the two directions). For $K_N$ the contribution
  of circuits of length $i$ to the sum is $i$ times the number of
  them, this gives Eq.~(\ref{eq:sigsum}).
\end{thebibliography}
\end{document}